\documentclass[%
amssymb,
]{article}

\usepackage{amsmath}
\usepackage{authblk}
\usepackage{blindtext}
\usepackage[a4paper, total={7in, 10in}]{geometry}
\usepackage{braket}
\usepackage{graphicx}% Include figure files
\usepackage{dcolumn}% Align table columns on decimal point
\usepackage{bm}% bold math
\usepackage{color,soul}

\begin{document}
%\preprint{APS/123-QED}

\newcommand{\notex}[1]{\textbf{\textcolor{red}{\hl{#1}}}}
\newcommand{\tb}[1]{\textcolor{blue}{#1}}
\newcommand{\tsr}[1]{\st{#1}}
\newcommand{\maximize}{\mathop{\mathrm{maximize}}}
\newcommand{\stt}{\mathrm{subject\ to}}
\setstcolor{red}
\def\eeps{\mathcal{E}}

\title{Robust optimal control for a systematic error in the control amplitude of transmon qubits}
% Achieving high fidelity for transmon qubits with uncertain control pulses 

\author[1]{Max Cykiert}
\author[1]{Eran Ginossar}
\affil[1]{
 University of Surrey, Advanced Technology Institute and School of Mathematics and Physics, Guildford, Surrey, GU2 7XH, United Kingdom.
}%
\affil[ ]{\textit {e.ginossar@surrey.ac.uk}}

\date{\today}% It is always \today, today,

\maketitle
\begin{abstract}

In the era of Noisy Intermediate-Scale Quantum computing as well as in error correcting circuits, physical qubits coherence time and high fidelity gates are essential to the functioning of quantum computers. In this paper, we demonstrate theoretically and experimentally, that pulses designed by optimization can be used to counteract the loss of fidelity due to a control amplitude error of the transmon qubit. We analyze the control landscape obtained by robust optimal control and find it to depend on the error range, namely the solutions can get trapped in the basin of attraction of sub-optimal solutions. Robust controls are found for different error values and are compared to an incoherent loss of fidelity mechanism due to a finite relaxation rate. The controls are tested on the IBMQ's qubit and found to demonstrate resilience against significant $\sim 10\%$ errors.

\end{abstract}

\twocolumn

%\tableofcontents
\section{Introduction}
Recently, the field of quantum information has been the subject of a great effervescence\cite{Gill2022,preskill2022}, fueled by hopes of a new type of computer that could outperform its fastest classical counterpart at specific tasks. Despite the fast development, many challenges remain at different levels of the implementation, from the fabrication, the calibration, the control and to the uses of such devices. While the field is transitioning from the academically interesting to the commercially viable, gaps in our understanding of optimal control remain.\\
Quantum computers require low error to perform reliably (as high as 1\% for the surface code\cite{Wang2011, Krinner2022}). However, the overhead cost of physical qubit per logical qubit can be in the thousands \cite{Fowler2012}. High fidelity gates will reduce the number of physical qubits per logical qubit. In order to implement quantum computations. Currently, single qubit gates on a superconducting qubit achieve fidelities exceeding 99.9\% for transmon type qubits \cite{Koch2007, Krinner2022,Jurcevic2021}, reaching the coherent limit.
The Hamiltonian parameters tend to drift in time due to the qubit itself or the control apparatus, requiring frequent re-calibration. 
Furthermore, the experimental method used to characterize the parameters of the system may themselves introduce some bias in the model leading to small mischaracterizations.

These drifts and mischaracterizations of parameters can lower the fidelity of quantum operations, but it has been demonstrated that optimization of the average fidelity  \cite{Dong2015,Dong2016,Carvalho2021,Kosut2022,Barnes2015,Zeng2018,Dong2021}, can reduce these effects. 
However, so far, pulses that maximize the worst-case fidelity for single qubit gates and their experimental realization have not been reported.
The worst-case optimization and average case optimization are compared in appendix \ref{AP:avvswc}.
Multiple techniques from the NMR community also exist correct different types of errors in the control and system \cite{Alway2007,Wimperis1994,Cummins} by using a composite pulse sequence. However, they are designed for two-level systems and are constrained by a rigid scheme where generally only a few different rotations are carried out at different driving angles, these methods are compared with those used in this work, as seen in the appendix \ref{AP:bb1}. The optimization in this paper takes advantage of optimal control strategies that have a less constrained format than the composite pulse sequences.

In this paper, we design, analyze, and implement a single qubit gate, $X_{\pi/2}$, for transmons. This gate is particularly convenient as it can generate any single qubit unitary in only a few operations. Furthermore, the constraints placed on the control promote obtaining solutions that are implementable in hardware. Finally, we compare these robust gates to a DRAG pulse\cite{DRAG}, a standard control technique that produces fast gates while preventing leakage.
We explore the control landscape, which maps the control variables to the fidelity, by using a local optimizer. This enables the comparison between the control landscape of the robust problem and the non-robust one. We also investigate how increasing the size of the error can change the control landscape.
We also analyze the stability of the fidelity against random perturbations on the control, this also allows the exploration of the vicinity of a solution. This enables the observation of some non-trivial sub-optimal trapping of the solutions during the optimization process as a function of both the error and the gate times.

We achieve robustness against a control amplitude error using the sequential convex programming optimization scheme. It samples the Hamiltonian at different error values and maximizes the worst-case fidelity. 
Optimizing the worst-case fidelity as opposed to the average fidelity guarantees a minimum performance in an error range  and the worst-case fidelity cannot be skewed by a higher fidelity at other points in the error sample. The two schemes are compared in the appendix \ref{AP:avvswc}.\\
In order for the pulses to be usable in practice, we impose stringent constraints on the control. The maximal and minimal amplitude off the control are limited and the bandwidth of the control signal is also limited, this is achieved using a Gaussian filter. 

Finally, the validity of the pulses is verified on the IBMQ's Armonk qubit. 
The pulses are calibrated by sweeping the amplitude on multiple channels and changing the phase of the signals. We use randomized benchmarking to measure the error per gate of the pulses and obtain a value of $3e\!-\!4$, and in the range of a $\pm10\%$ artificial error on the control amplitude the error per gate is no worse than $5e\!-\!4$.

\section{Qubit model and control}
The transmon qubit is modeled as a three-level system, with the first two levels representing the computational subspace and the third level the leakage level.
The Hamiltonian in the rotating frame after making the rotating wave approximation is (with $\hbar=1$)
\begin{equation}
    H(t)=\sum_{j=1,2}^N\left[ \delta_j \Pi_j + \frac{\eeps^x(t)}{2}\lambda_j \sigma^x_{j-1,j}+ \frac{\eeps^y(t)}{2}\lambda_j \sigma^y_{j-1,j} \right].
    \label{eq:hami}
\end{equation}
Where $\Pi_j=\ket{j}\bra{j}$, $\sigma^x_{k,j}=\ket{k}\bra{j}+\ket{j}\bra{k}$ and $\sigma^y_{k,j}=i\big(\ket{k}\bra{j}-\ket{j}\bra{k}\big)$. The detuning of the transitions to the drive frequency are $\delta_1=\omega_1-\omega_d$ and $\delta_2=\Delta+2\delta_1$ with $\omega_1$ being the qubit 0-1 energy difference and $\Delta$ being the anharmonicity. Also, where $\lambda_i$ is the maximum Rabi-rate achievable between the i\textsuperscript{th}-1 and i\textsuperscript{th} level. It is worth noting that this value of the rabi-rate will limit the minimum speed at which a gate can be carried out .
%\subsection{Constraints on control}
%\label{subsec:constr}
The control signal is quadrature amplitude modulated on a drive frequency  $c(t)=\eeps^x(t) cos(\omega_d t) + \eeps^y(t)sin(\omega_d t)$, where $\eeps^x$ is the in-phase component and $\eeps^y$ is the quadrature component. The time evolution of the qubit is considered during $0<t<T$, and T is the gate time. We will use the Hamiltonian to be piece-wise constant (PWC) as an ansatz and the $H_i$ notation to refer to the i\textsuperscript{th} piece of the Hamiltonian. 
The values of the parameters are based on the IBM qubit Armonk (see table \ref{tab:table1}). 
\begin{table}[b]%The best place to locate the table environment is directly after its first reference in the text
\begin{center}
\begin{tabular}{|c | c|}
\hline
Parameter & Value\\
\hline
$\omega_{01}/2\pi$ & 4.974 GHz \\ 
$\Delta /2\pi$  & 345 MHz \\
$\lambda_1/2\pi $ & 15 MHz\\
$\lambda_2/ 2 \pi$ & 15 MHz\\
$\omega_b/2\pi$ & 24 MHz\\
\hline
\end{tabular}
\end{center}
\caption{\label{tab:table1}%
The parameters used for the optimization. Note that all the pulses are on resonance ($\delta=0$), including the DRAG pulses.
}
\end{table}
$\lambda_2$ was determined while on resonance with $\omega_{01}$ by a Rabi-like experiment between the first and second levels. Furthermore, no correction was applied due to a finite bandwidth of the signal generator, which would make the $\lambda_2<\sqrt{2} \lambda_1$. We impose a linear convex constraint on the maximum control amplitude, which is, of course, always the case in experiments,
$\left|\eeps_{x,y}\right|\leq 1/\sqrt{2} $ which satisfies $\sqrt{\eeps_x^2+\eeps_y^2}\leq1$.  
%\subsection{Filtering And Slew Rate}
Arbitrary wave generators (AWG) have a finite bandwidth that cannot reproduce faithfully any PWC controls that produce high fidelity operations. Therefore, if no constraints are imposed on the control signal, some of the solutions obtained by the optimization may be unimplementable in physical hardware. Hence a control must not have large frequency requirements nor big swings in amplitude from minimum to maximum in a few time bins. We filter the controls to avoid high frequency components by choosing a Gaussian filter with a small bandwidth, $\omega_b$, as an internal parameter to the optimizer (see table \ref{tab:table1}). This small bandwidth is more stringent than typical experimental apparatus' bandwidth. In this system, as the gate time is significantly longer than the inverse of the bandwidth  $T > 2\pi/\omega_b$ , the control has sufficient time to reach the maximum amplitude during the gate duration. In addition, the narrow bandwidth of the resulting signals avoids the need to pre-distort the pulse based on the finite bandwidth  frequency filter of the AWG's and instead tune the pulse's amplitude.
The filtering effectively maps the control variables to the control signal, where the two may look vastly different from each other.
The control can be filtered a second time using an inverse filter function to compensate for the experimental apparatus. However, doing this would only have a slight effect on the signal as the bandwidth of the pulse is much smaller than the hardware's. 
This kind of filtering effectively enforces the smoothness of the pulse.
\begin{equation}
    \eeps(t)=\mathcal{F}^{-1}\bigg[G(\omega)\mathcal{F}[c(t)]\bigg],    
\end{equation}
where $\mathcal{F}$ is the Fourier transform of the $c(t)$ signal, and G is the filter function.
A linear transform can approximate this filtering\cite{Motzoi2011}:
\begin{equation}
    \eeps^{x,y}=Mc^{x,y}
    \label{eqn:LT}
\end{equation}
where $M$ is a matrix that is not necessarily square.  An advantage to expressing the filtering as a linear function is that the gradient of the fidelity with respect to the control variables is simply multiplied by the same transfer matrix $M$, see equation \ref{eqn:pert}.\\From here on out, we will choose the number of rows of $M$ to be $4\left|c_x\right|$, meaning that each control variable is subdivided into four smaller bins. The control can be further smoothed by simply changing the number of rows in M, and this can be useful if the solution's signal samples have to be matched with the sampling rate of the arbitrary wave generator. 
The control vector, $\eeps_{x,y}$  is padded with $n_0$ zeros at the beginning and the end of the control vector to ensure that the control signal starts and ends at 0 amplitude, as Eq. \ref{eqn:LT} does not guarantee it. Enforcing this condition prevents high frequency components to the control due to the sudden start and end of the control signal. 
The amount of zeros used in the padding is determined by considering a discrete Heaviside step function having its step at index $n_0+1$ such that the first entry of $\eeps_x$ is less than $0.1\%$ of the maximum amplitude.
\begin{equation}
    n_0=\min_i s.t. \sum^N_{j=i} M_{1j} < 0.001
\end{equation}
Furthermore, in order to limit the controls variable from changing too significantly before filtering, a slew constraint is set on the control vector such that $|c_i-c_{i+1}|<1$.

\subsection{Error Model and Robustness}
The qubits and the control apparatus itself can suffer from many kinds of distortions \cite{VanDijk2019}, and robust optimization, in general, can mitigate some of the effects of these non-idealities. 
In this paper, we consider an error $\eta$ that is time-independent, for the duration of the gate, of the coupling between adjacent energy levels $\eeps^{x,y} \rightarrow (1\pm\eta)\eeps^{x,y} $. 
During the optimization, the Hamiltonian is sampled three times at different error values, $\eta\in\{-\eta, 0, +\eta\} $. It is essential to keep the zero-error Hamiltonian, $\eta_i=0$, to avoid a solution that has a worse fidelity at the most probable point. We note that at a small error range, it might not be necessary to sample the zero error point \cite{Le2021}, and at a large error range, it may be worth sampling more points\cite{Dong2015}.
When repeated measurements of the Rabi rate between the 0-1 energy levels and the 1-2 energy levels were carried out on the IBMQ Armonk qubit, we observed a standard deviation of around 0.3 percent and 1 percent. A fluctuation on the parameter $\omega_{01}/2\pi$ of a much smaller order, which was measured to be $\sim$10 KHz inline with a few KHz \cite{Burnett2019} or $\sim$100 KHz in \cite{Schlor1905}. The effect of this error in the qubit frequency is much smaller than the error in the Rabi rate; therefore, it will have a minor effect in the timescale of the gate. In fact, at a second order approximation the error due to the amplitude is about 1000 times greater than the one due to the detuning. The value of the detuning error which gives an error equal to the error due to the amplitude error is around 335 kHz (see appendix \ref{AP:error}). Nevertheless, the techniques used in this work can also be used to prevent the fidelity loss due to a frequency error (see appendix \ref{AP:freqRob}).
\begin{figure*}
  \centering
  \includegraphics[width=0.48\linewidth]{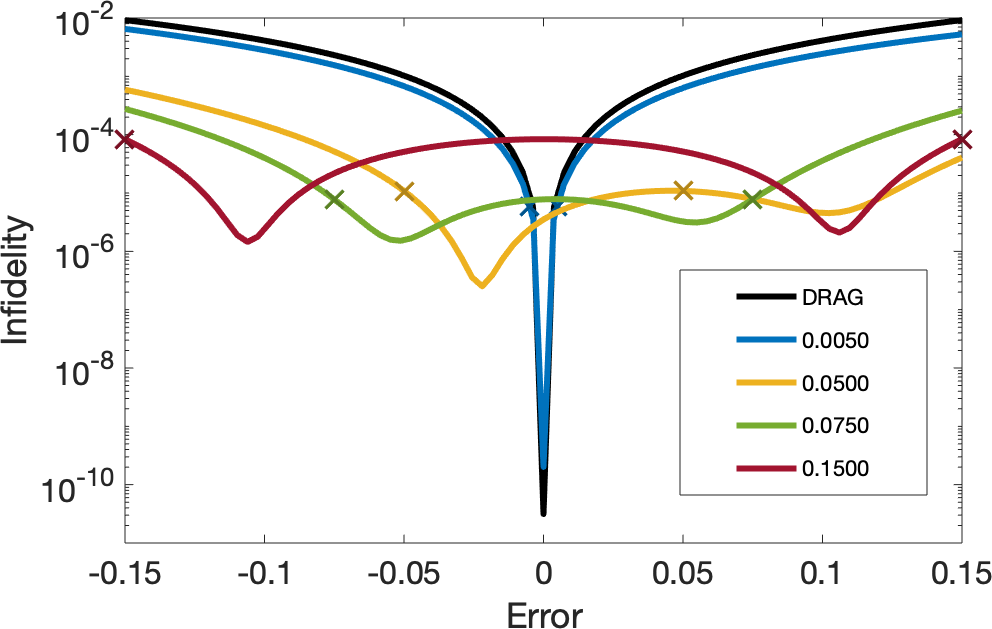}
  \hspace{3pt}
    \includegraphics[width=0.48\linewidth]{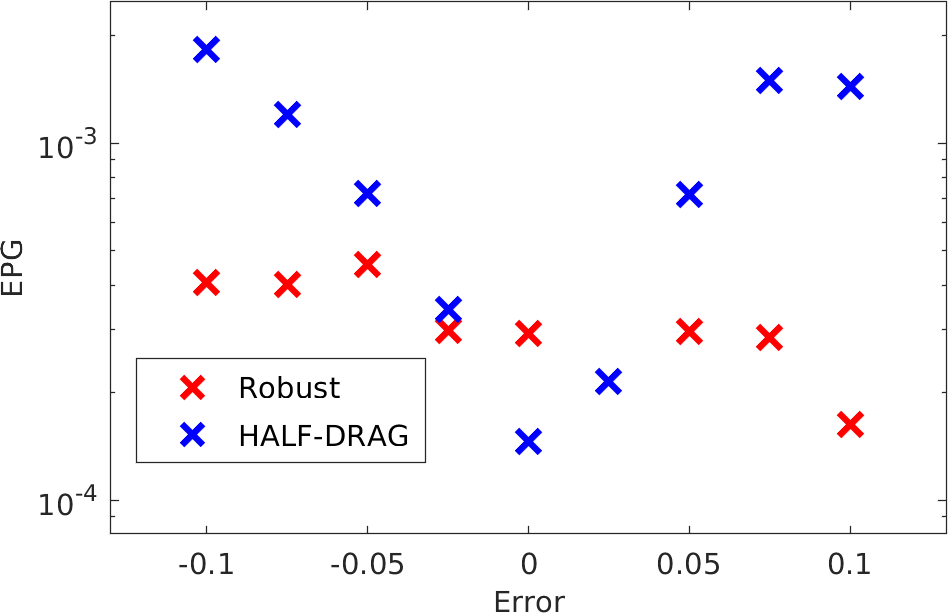}
    	\caption{Left panel: Values of the infidelity as a function of 1 plus the error on the control amplitude for an optimized pulse. The pulse duration is T=150ns, and there is 50 control variable per channel. The colored lines are optimized robustly and the black line is a DRAG pulse optimized non-robustly. The crosses represent the error for which the pulses were optimized, and the zero error sample cross is not included. Note that the values outside the crosses are outside of the robust range.
    	Right panel: Experimental randomized benchmarking experiment for a robust pulse and a half-DRAG pulse, as a function of an artificial error on the control amplitude, realized a hundred hours after the initial calibration. The gate time of the DRAG pulse is 72ns, and the robust pulse is 130ns.}
    	  %\label{fig:compa}
    \label{fig:RB-timeStab}
    %SHOULD IT BE T_1/T_g
\end{figure*}

%\subsection{Non-robust solution, DRAG}
% \begin{align}
%     \eeps^x(t)=\eeps_\pi + \frac{(\lambda^2-4)\eeps_\pi^3}{8\Delta^2}-\frac{(13\lambda^4-76\lambda^2+112)\eeps_\pi^5}{128\Delta^4},\\
%     \eeps^y(t)=-\frac{\dot\eeps_\pi}{\Delta}+\frac{33(\lambda^2-2)\dot\eeps_\pi\eeps_\pi^2}{24\Delta^3},\\
%     \delta_1=\frac{(\lambda^2-4)\eeps_\pi^2}{4\Delta}-\frac{(\lambda^4-7\lambda^2+12)\eeps_\pi^4}{16\Delta^3}.
% \end{align}
DRAG is a standard control technique that effectively reduces leakage to the third qubit level in fast pulses. 
We will use DRAG pulses to compare the robustness of the pulses obtained in this paper.
Another benefit of this method is that it has an analytical form where at first order, the quadrature component of the pulse is proportional to the derivative of the in-phase control envelope. DRAG is often used in experiments \cite{Kandala2019}, and a simplified version exists, half-DRAG, that can compensate for a phase error in the control by scaling the derivative \cite{Lucero2010,Rahamim2017}.
% \begin{align}
%     \eeps^x(t)=\eeps_\pi + \frac{(\lambda^2-4)\eeps_\pi^3}{8\Delta^2}-\frac{(13\lambda^4-76\lambda^2+112)\eeps_\pi^5}{128\Delta^4},\\
%     \eeps^y(t)=-\frac{\dot\eeps_\pi}{\Delta}+\frac{33(\lambda^2-2)\dot\eeps_\pi\eeps_\pi^2}{24\Delta^3},\\
%     \delta_1=\frac{(\lambda^2-4)\eeps_\pi^2}{4\Delta}-\frac{(\lambda^4-7\lambda^2+12)\eeps_\pi^4}{16\Delta^3}.
% \end{align}
The DRAG pulse achieves the highest fidelity at zero error. However, the DRAG pulse is not innately robust against an error in the control amplitude, which can be seen from the fidelity's rapid decay proportional to $\eta^2$.
Under certain conditions, this decay can be illustrated by an error on the control's amplitude for a qubit (see appendix \ref{AP:error}).
However, DRAG shows robustness against an error on the coupling to the leakage level, $\lambda_2$ \cite{Motzoi2013, Theis2018}.
The robust pulses are compared to non-robust pulses which are generated using the analytical DRAG controls as the starting point of a non-robust robust optimization, in order to reach unit fidelity up to numerical precision.\\
The DRAG pulses are optimized by the same method as described above with the only difference being that the Hamiltonian is sampled only once at zero error, and we note that after the optimization the pulse reaches unit fidelity (up to numerical error). While the half-DRAG pulses are tuned by changing the amplitude of $\eeps^x$ and $\eeps^y$.\\
After the optimization, the DRAG controls retained the shape that they started with and kept their smooth and continuous features.

\section{Optimization techniques}
\subsection{Objective function}

We will consider two ways to calculate the system's evolution: By integrating either the time-dependent Schrödinger equation or a Lindblad-Master \cite{Navarrete-Benlloch2015} equation. The optimization will be carried out using the former, and the incoherent effects and the gate duration will be understood using the latter. 

The time evolution of a PWC Hamiltonian generates a unitary matrix $U$:
\begin{equation}
U(T)=\mathcal{T}\prod_{j=1}^{N_t}\exp(-i H_j dt)
\label{eq:evo}
\end{equation}
where $\mathcal{T}$ is the time ordering operator, $N_t$ is the number of time steps and $dt$ is the duration of a piece-wise constant pulse, and $H_j$ is the Hamiltonian at the j\textsuperscript{th} time step.

The evolution of the qubit is not in practice unitary; the incoherent effects of the system are not negligible and can be described by a Lindblad-Master equation:
\begin{equation}
    \dot\rho=-i[H,\rho]+\sum_{j=1,2}\bigg(\frac{D[\sigma_j^-]}{T_1^j}+\frac{D[\Pi_j]}{T_\phi}\bigg)
    \label{eqn:master}
\end{equation}
where $D$ is the super-operator $D[\sigma]=D[\sigma,\rho]=\sigma \rho \sigma^\dagger - \{ \sigma^\dagger\sigma,\rho \} /2 $. We will consider cases where $T^1_1=T^2_1$ and $T_\phi= \infty$ to represent qualitatively incoherent effects.

In the case of a unitary evolution, the evolution of the density matrix is simply $\rho(T)=U\rho U^\dagger$. 

The target unitary in the paper will be a three-level $X_{\pi/2}$, with the identity in the third level. This target gate together with a parametric Z-gate forms a universal gate set and can be used to generate any SU(2) gate using exactly three virtual-Z gates and two $X_{\pi/2}$\cite{McKay2017}. Computing only this gate reduces the computational cost of calculating a universal gate set containing 2 or 3 gates \cite{Nielsen2010} and hastens the tuning process.

A performance indicator is needed to guide the optimization, i.e., an objective function, and multiple candidates exist for this task. This objective function would benefit from having a simple form and depending only on the time evolution operator. A natural candidate could be the state overlap. It has one of the simplest form, making an objective function, $\bra{\psi}\ket{\psi_T}$, this, however, outputs a complex number, so this could be squared and would provide the quantum state fidelity, $\left|\bra{\psi_0}U_T^\dagger U\ket{\psi_0}\right|^2$. However, this quantity depends on the input state $\ket{\psi}$, and may give the spurious value of unity for some initial states despite $U$ being sub-optimal\footnote{If $M=U^\dagger_TU$, M is unitary and there exists an eigenvector $\ket{\psi}$ such that $M\ket{\psi}=e^{i\phi}\ket{\psi}$, giving a fidelity of 1 for this eigenvector}. On the other hand the gate fidelity $F_1$\cite{Magesan2011,Fortunato2002}, is state independent:
\begin{equation}
F_1(U,U_T)=|tr(U^\dagger U_T)|^2/{n^2}
\end{equation}
Where U is the obtained unitary matrix and $U_T$ is the target unitary.
It has the property $0\leq F_1(U_T,U)\leq 1$ where $F_1(U_T,U)=1\iff U_t=U$, which makes it practical as an objective function, furthermore this quantity can be related to the average fidelity on the Hilbert space \cite{Pedersen2007}. 
This definition of the gate fidelity is state independent and only depends on the implementation of the unitary matrix $U$, but it is sensitive to the relative phase in the third level. For instance, a perfect X gate in the qubit subspace could still have a phase of $\pi$ in the third level, which would lower the fidelity to $F=1/9$. However, this lowering of the fidelity would not be relevant in the context of qubit's operations\cite{Rebentrost2009}.

In order to fix this last issue, another quantity derived from the gate fidelity is used, where the fidelity is taken only on the qubit subspace:   
\begin{equation}
  F_2(U,U_T)=|\sum_{j=0,1}(U^\dagger U_T)_{jj}|^2/{2^2},
\end{equation}
which still penalizes leakage but is not sensitive to a phase in the third level.
While $F_2$ measures how well the unitary evolution has been achieved,  it does not correspond to a physical quantity and is regrettably not easily measurable, which is a desirable feature\cite{Goerz2017}; 
%This quantity also has the advantage of not having fidelity losses due to an irrelevant phase in the third level, which is mostly unpopulated.  

When the qubit evolves according to Eq. \ref{eqn:master}, another definition for the fidelity will be used one that depends on the initial state. However, when that fidelity is averaged over 6 poles of the Bloch-Sphere it is equal to the average fidelity of all the qubit state\cite{Bowdrey2002}.
The average fidelity can therefore be calculated for the density matrix, $\rho$, after either the evolution from a time dependent Schrödinger equation or the Master equation evolution in Eq. \ref{eqn:master}:
\begin{equation}
    F_A=\frac{1}{6}\sum_{j=\pm x\pm y\pm z}tr(U_T\rho_j(0)U^\dagger_T \rho_j(T)),
    \label{eqn:FidelityAverage}
\end{equation}
In the context of pure states, $F_A$ corresponds to the average probability over all the qubit input states of measuring the qubit in the state $U_T\ket{\psi}$.

In experiments, the measurement of the qubit state has a lower fidelity than the fidelity of the gates, therefore in experiments, the measure of success is often obtained from randomized benchmarking, a technique that fits a decay of the qubit state after a sequence of Clifford operations which should ultimately leave the state of the qubit unchanged. The error per gate\cite{Proctor2017} (EPG) is extracted from this decay and quantifies the performance of the gate (see section \ref{sec:experimental}).

We will use $F_2$ as the objective function during the optimization as it is marginally faster to compute than $F_A$, and all subsequent values quoted for the fidelity and infidelity will be from $F_A$. 
We will define the infidelity as $1-F$, which is more convenient when dealing with numbers close to 1.

\subsection{Sequential Convex Programming}
In optimal control theory, the control landscape is mapping the control variables to the fidelity\cite{Brif2010}, and the goal of the control landscape exploration is to find a local or global maximum.
The quantum control landscape is considered to have a lot of favorability to optimization because most critical points are almost always saddles \cite{Brif2010,Chakrabarti2007}, implying that the solution reached following a gradient ascent will lead to a global maximum. This make the use of local optimizers, i.e. gradient based optimizers, very well suited.
The assumptions here are that the control is unconstrained in time and amplitude, that it has enough resources.
These conditions can, in fact, be relaxed to have finite resources and still provide excellent answers.
However, under even more relaxed conditions control landscapes still having sufficient resources, can become filled with traps, this can happen when the number of control variables is reduced, the gate time is reduced or constraints on the limit of the amplitude of the control is reduced\cite{Moore2012,MooreTibbetts2012,Zhdanov2015,Donovan2015}.

In this paper, we use Sequential Convex Programming \cite{Kosut2013, Allen2017, Allen2019} (SCP) to obtain robust solutions to control transmon qubits which have more than two levels. SCP is a variant of a gradient ascent method,
it allows the simultaneous exploration of multiple control landscapes, where each landscape is a sample at different error values.
The fidelity is improved by calculating the gradient of different landscapes and then solving the linear program problem that finds an update that increases the fidelity of every control landscape.
SCP is different from other optimization scheme such as GRAPE\cite{Khaneja2005}, which maximizes the fidelity by updating the control proportionally to the gradient of a singular control landscape. On the other hand, SCP maximises the worst-case fidelity:
\begin{equation}
    \label{eqn:pert}
    \begin{array}{ll}
        \underset{x}{\maximize} \quad\underset{i}{\min}(F^i +M\nabla F^i M x)\\
        \stt \quad -\lambda \preceq x \preceq \lambda
    \end{array}
\end{equation}
where $x$ is the update on the control vector $c\rightarrow c+x$, $\lambda$ is the trust region, and $M$ is a linear transfer function that has been defined previously. Here, the $F^i$ is the fidelity function associated with the i\textsuperscript{th} Hamiltonian sample, $H^i$. If there is no transfer function, $M$ can be set to 1.
After every update $x$, SCP verifies if the fidelity increases and if the fidelity on all landscapes has improved, the update is accepted, and the trust region $\lambda$ is increased by a factor of $a_{incr}$. Otherwise, the trust region is decreased by a factor of $a_{decr}$ and then this optimization step is repeated. The adaptive change in the trust region:
%SCP solves the linearised fidelity programming problem. 

\begin{eqnarray}
IF &&\quad \forall i\ ,\ F(H^i,c+x)\geq F(H^i,c)\nonumber\\
THEN &&\quad c\rightarrow c+x, \lambda \rightarrow \lambda  a_{incr}\\
ELSE && \quad\lambda \rightarrow \lambda  a_{decr}\nonumber
\end{eqnarray}
This optimization problem is solved in MATLAB, using the YALMIP \cite{Lofberg2004} interface and the quadprog solver, which was found to give the highest fidelity compared to others.  

This optimization process is repeated until one convergence condition is met: a maximum fidelity threshold is attained, the trust region is below a specific value, or the change in fidelity between n consecutive iterations is below a threshold. The difference in fidelity between successful iterations might vary considerably, so we average this difference over the last ten iterations to avoid an early termination.
The stopping conditions used in the optimization are shown in table \ref{tab:table2}.
\begin{table}[bh!]%The best place to locate the table environment is directly after its first reference in text
\begin{center}
\begin{tabular}{| c | c |}
\hline
Parameter & Value\\
\hline
Number of starting points & 5000\\
Number of iterations & 1e4\\
\textrm{Stopping criteria}&\\
Fidelity & 1 \\ 
Fidelity difference between iterations  & 1e-10 \\
Trust Region  & 1e-9\\
\hline
\end{tabular}
\end{center}
\caption{\label{tab:table2}%
The parameters used for the optimization. The robust pulses are on resonance ($\delta=0$), while the DRAG pulses are constantly detuned in time.
}
\end{table}

\section{Results}
Firstly, the worst-case fidelity in the region is rarely found to be worse than the fidelities sampled at the three points during the optimization, and in the rare cases when it is, the difference is insignificant, implying that sampling three times is adequate.\\ 
When considering a unitary time evolution, the three most important parameters are the gate time, the number of control variables, and the error range. Generally, increasing the number of control variables will allow more complex dynamics, which can refocus different errors on the evolution. However, when a linear transfer function filters the control, as described by Eq. \ref{eqn:LT}, the control variables induce a smaller change in the control signal amplitude as compared to the unfiltered case, the signal at a specific time becomes dependent on to the variables at other time points. For example, the pulse in section \ref{sec:experimental} has an amplitude that is a linear combination of the control variables, and a control variable will only contribute 32\% to the amplitude of the control signal at the corresponding time step of the signal. This exemplifies why a single control variable in an optimization with many control variables and a filter function will be less capable of affecting the resulting signal and the qubit dynamics. Therefore, a large number of control variables in a filtered signal will be less apt at improving the fidelity but will needlessly increase the optimization time.
\begin{figure}[h!]
    \centering
    \includegraphics[width=0.8\linewidth]{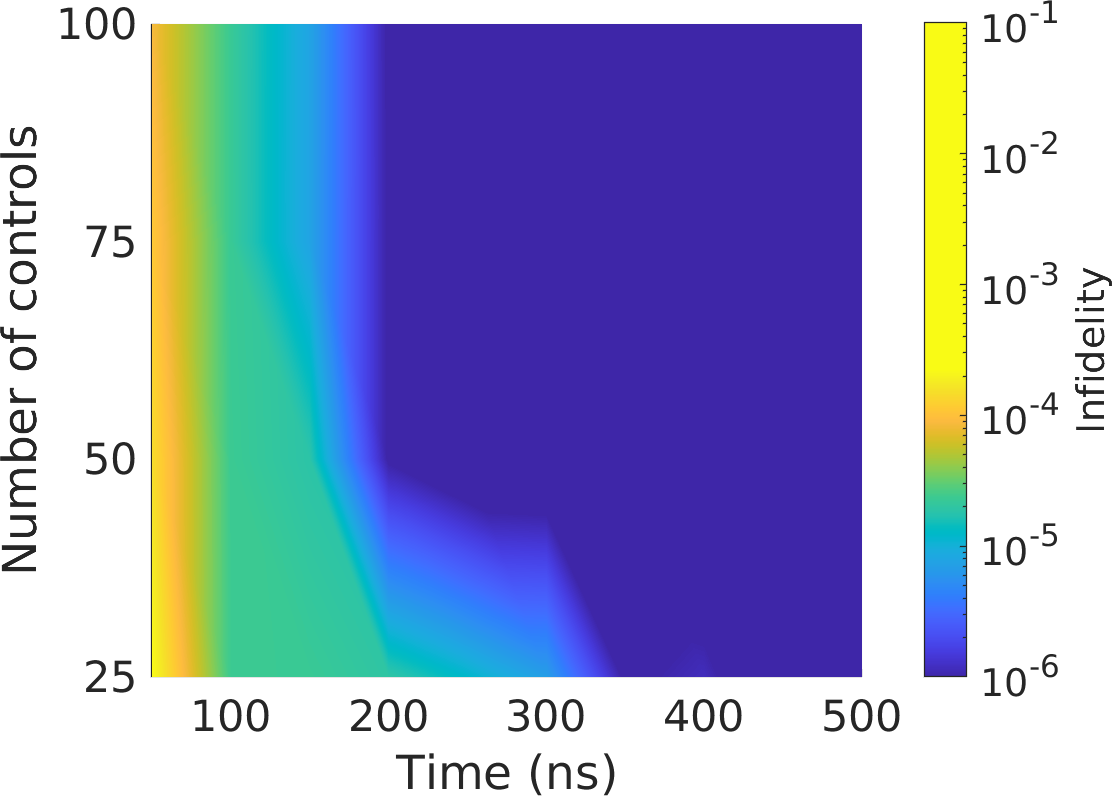}
    \caption{Infidelity as a function of the number of control variables and gate times. The Z-axis is the worst-case infidelity at $\eta=\{0,\pm5\%\}$ for a unitary evolution of the robust control. The map is interpolated from $t = 50ns$ by increments of 50ns, and the number of controls ranges from 25 to 100 by increments of 25 controls per channel.}
    \label{fig:heatmap}
\end{figure}

At gate times slightly shorter than around 120ns, solutions that have a high fidelity at the extrema of the error range are hard to find. The fidelity is just maximized in the middle of the range and the fidelity across the error range is very similar to the non-robust solution's fidelity. However, the fidelity also increases as the gate time increases from these short gate times.

\begin{figure}[h!]
     \includegraphics[width=0.55\linewidth]{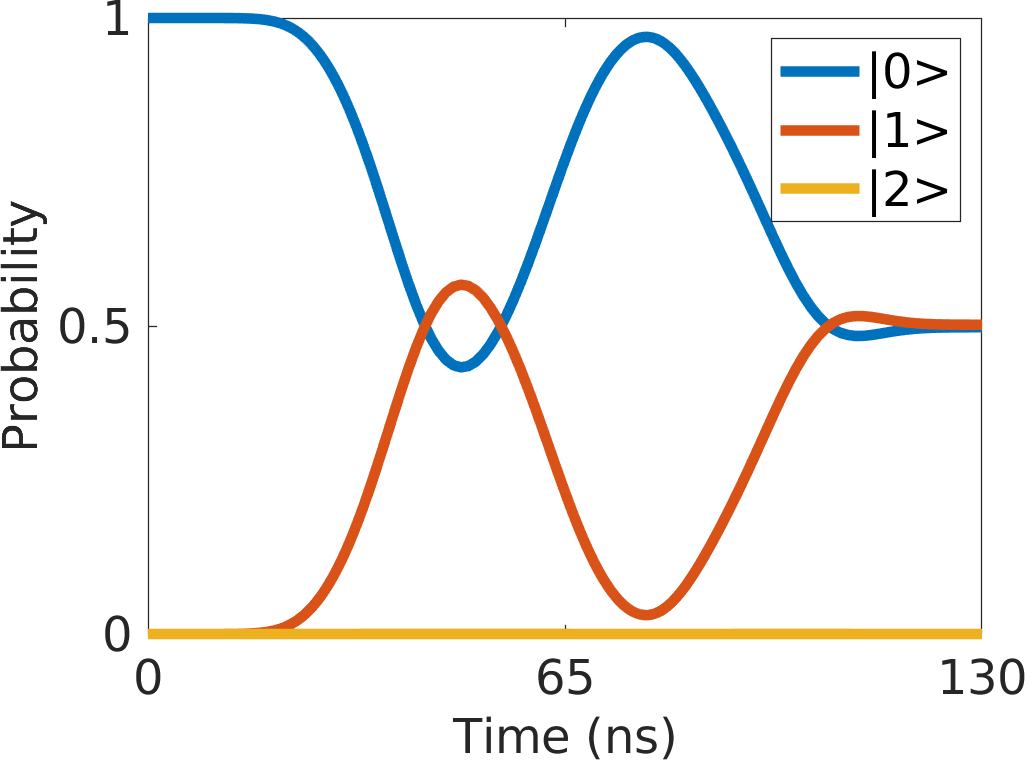}
     \label{fig:population}
     \quad
     \includegraphics[width=0.35\linewidth]{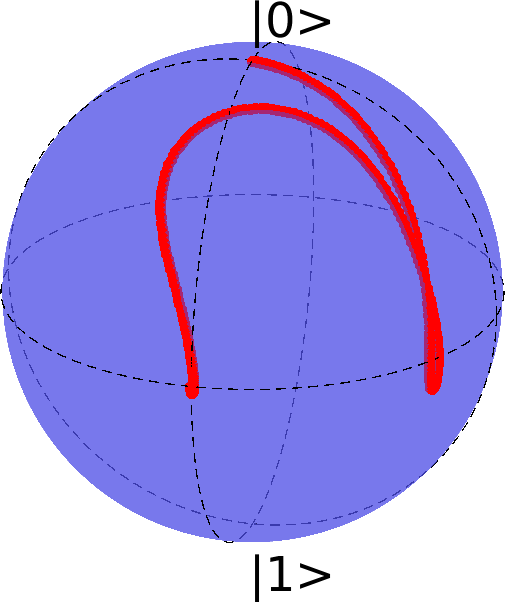}
     \label{fig:BLOCH}
    \caption{Time evolution of the control. Left panel: Probability of the control from figure \ref{fig:control} with $\eta=0.075$, the populations of the first three levels in blue, orange, and yellow, respectively. The maximum leakage is $<$ 1e-4. Right panel: Bloch sphere representation of the same trajectory starting in the ground state.}
    \label{fig:populationANDBloch}
\end{figure}

The number of control variables impacts the fidelity that can be achieved, especially at low timescales $150 ns<T<300ns$ where the number of control variables increases from 25 to 75, the fidelity also increases by an order of magnitude, see Fig.~\ref{fig:heatmap}.
At longer gate times, $T>300ns$, the worst-case fidelity plateaus, and no more gains in fidelity are observed.
The controls obtained from the optimization are smooth and start and end at zero amplitude (see Fig.~\ref{fig:control}).
The trajectories for the robust control are more convoluted than the DRAG trajectory, see figure \ref{fig:populationANDBloch}, the angle of driving, $\theta(t)=\mathrm{atan}(\eeps^y(t)/\eeps^x(t))$, changes throughout the pulse. The most significant amount of leakage obtained during the pulses in all the optimization from figure \ref{fig:heatmap} is less than $0.15\%$, and the median largest leakage is less than $0.0069\%$.

In practice, it is a balancing act between finding a robust pulse that is long enough to bring benefits in terms of worst-case fidelity across the error range and a gate time that is short enough to not limit the fidelity due to the incoherent effects, see section \ref{sec:competingLosses}.

At short gate times, much less than around $<$120ns, robust solutions that have a significantly higher fidelity at the extrema of the error range are impossible to find. There are few benefits to using robustly optimized controls at these gate times as they perform similarly to a non-robustly optimized control. When the gate time is short, the fidelity is maximized in the middle of the range, similarly to a non-robust optimization.
However, as the gate time increases, the fidelity will also increase. The number of control variables impacts the fidelity that can be achieved, especially at low timescales $150<T<300$ where more control variables increase the fidelity by an order of magnitude see Fig.~\ref{fig:heatmap}.

\begin{figure}[h!]
    \centering
    \includegraphics[width=0.7\linewidth]{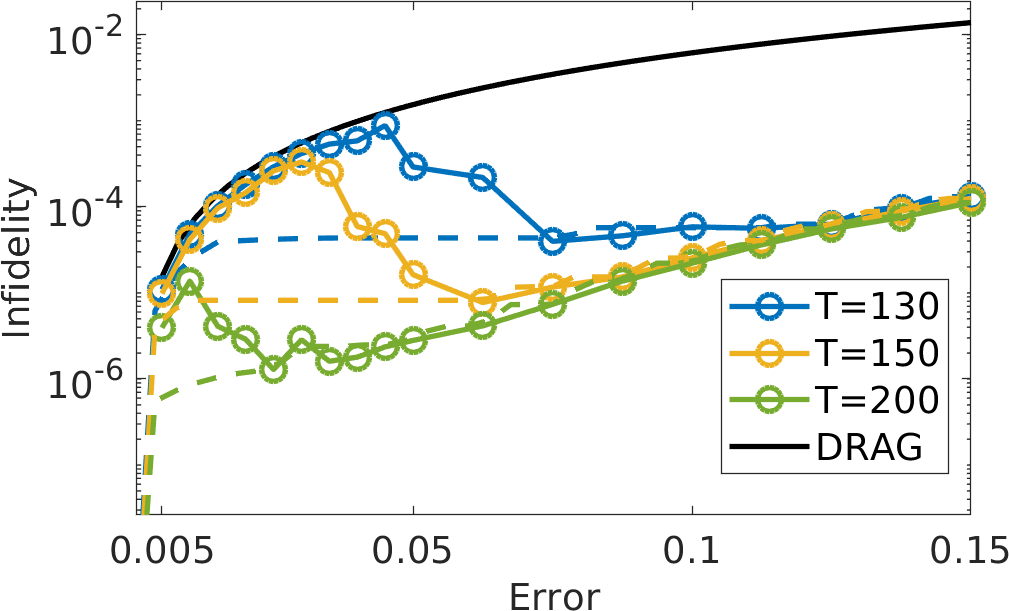}
    \caption{Dependence of the fidelity on the error. Comparing the worst-case infidelity inside the range  $[-\eta, \eta]$, which was sampled 41 times. The controls were optimized for an error of $\eta$. The controls had 50 variables per channel. Every point is the highest fidelity obtained out of 5000 starting points. The dashed lines are the solutions with the highest fidelity in the sub-region $[-\eta, \eta]$ out of all the other solutions (shown by the markers).}
    \label{fig:robustBenefit}
\end{figure}

The worst-case fidelity depends also on the error: the maximum worst-case fidelity found using robust optimization was for an error value of $\eta=0.035$ and is 1e-6. 
Counter-intuitively, the optimization found controls with higher worst-case fidelity when the error sample was larger.
Counter-intuitively, in many cases, optimizing for a larger error range yielded higher fidelities.
The dashed lines in figure \ref{fig:robustBenefit} represent illustrate this: the highest fidelity found for an error value and the solid line are the fidelity found when optimized for that error.
In fact, the most considerable difference in fidelity between the best available fidelity for an error range and the fidelity of the controls optimized for that error can be as high as 1.5 orders of magnitude (see figure \ref{fig:robustBenefit}, the difference between the dashed and solid yellow lines). However, the best fidelity at an error range and the fidelity optimized for that error eventually converge.
When the error is small, see for instance, the points in figure \ref{fig:robustBenefit} which are on the DRAG curve, the optimization gets trapped on lower worst-case fidelity. As a result, these solutions do not exhibit robustness and have a similar profile to a non-robust solution. This indicates that the basin of attraction of the sub-optimal controls is larger than that of a more optimal one and that the effect of the error changes the control landscape too little for the optimization to find the robust solutions.
The fidelities can be sorted in three areas based on the value of the error, for low errors, the robust pulses perform similarly to non-robust pulses, their fidelity match the DRAG fidelity profile. While for large errors, the pulses perform robustly, and in between the low and large error zones, the fidelity transits from the non-robust fidelities to the robust ones. Also, trapping occurs most often in the the non-robust and transient areas (see Appendix \ref{sec:Hrdness},\ref{sec:pert}). 
In figure \ref{fig:robustBenefit}, the dashed line being below the solid line directly implies that the optimization converges on sub-optimal solutions; that there are traps in the landscape.
The points which are on the DRAG infidelity line are unable to find robust solutions but rather get trapped on solutions which are in the vicinity of non-robust solution. At low error range the basin of attraction of non-robust solutions is larger than the robust one. At higher gate times, the fidelity increases and the region where the traps occur decreases.

%that a control optimized for a greater error and produced a higher fidelity in the sub-region of the error is used instead of the one it was optimized for (solid line).
%The error range is sampled 83 times, while it was sampled three times during the optimization.

We note that the fidelity trapping due to low gate time or low error range can be remedied by introducing cycles of perturbations and re-optimizations (see Appendix \ref{sec:pert}), this is effective and in some cases increased the fidelity by one order of magnitude.
The low fidelity due to low time resources can be remedied by introducing cycles of perturbations (see Appendix \ref{sec:pert}) and re-optimization (see the right panel of Fig.~\ref{fig:fractions}) on the controls keeping only the best results for the next run. Using this method with three such cycles, we could increase the fidelity by more than one order of magnitude, untrapping the control's optimization.
%\note{needs updating}The basin of attraction can be further quantified by perturbing the control and seeing if the perturbed control return to the original fidelity. The controls with the ten highest fidelities of the 250ns with 2*50 controls were perturbed while respecting the constraints described in the first part of the paper. These constraints could tightly bind the control, so the perturbation process was repeated to generate larger perturbations.

Comparatively, optimizations on a single fidelity landscape (non-robustly) find a global maximum with unit fidelity if the Hamiltonian is controllable and there are enough resources: time and no constraints on the drive amplitude\cite{Russell2017}. Even with the control having a limited amplitude, high fidelity will still be achieved\cite{Moore2012}.
In contrast, all other parameters being equal, the landscape is trapped in a region $0<\eta<0.025$, or for a short gate time, longer than the time needed to have a non-robust gate\cite{Ashhab2012}, makes the optimization prone to trapping. Including the robustness from Eq. \ref{eqn:pert} changes the control landscape significantly compared to the non-robust case, the probabilities of finding good solutions becomes lower (see Appendix \ref{sec:Hrdness}).

\subsection{Competing fidelity losses processes}
\label{sec:competingLosses}
\begin{figure}
    \centering
    \includegraphics[width=0.49\linewidth]{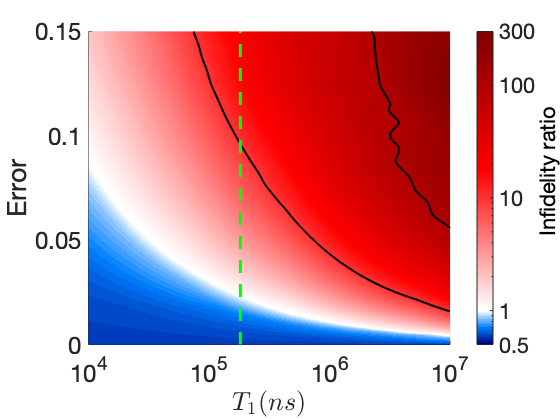}
    \includegraphics[width=0.49\linewidth]{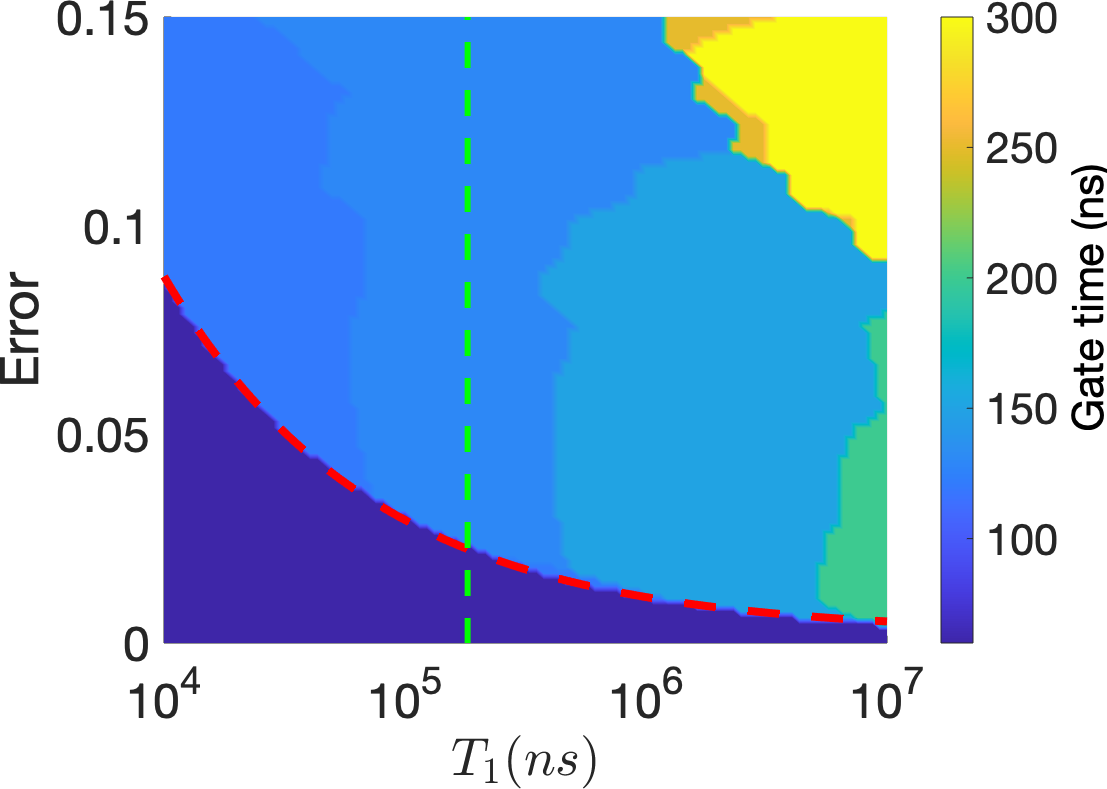}
    \caption{Left panel: highest fidelity achievable at different values of $T_1$ and of the control amplitude error. ratio of the infidelities of a 60 ns numerically optimized pulse and the best performing robust pulse for different gate times. The contour lines are at ratios of 10 and 100. The dashed line is the value of the qubit's $T_1$, which is 182$\mu s$.
    The dashed red line is the boundary between when robust pulses achieve a higher fidelity (above) compared to non-robust pulses (below). Right panel: gate time that achieve the highest fidelity, the robust gate time considered are 120, 130, 150, 200, 250, 300 ns. The non-robust pulse duration is 60 ns long.}
    \label{fig:eVST1}
    %SHOULD IT BE T_1/T_g
\end{figure}
In practice, robust pulses are only interesting if they can provide advantages over their non-robust counterparts. This will depend on the classical controls and the qubits. 
The added gate duration of a robust pulse over a non-robust comes as a detriment to the fidelity during an incoherent evolution and is in direct competition with the fidelity gains due to the robustness. Therefore the robust control must overcome the added duration, around a factor of two longer than the non-robust pulse before they can outperform non-robust solutions such as DRAG.
We next consider the effect of decoherence on the performance of the gate. For simplicity we include only the relaxation process ($T_1$) and neglect the effects of pure dephasing. This represents the majority of transmon devices which tend to have very small pure dephasing rate compared to the decoherence induced dephasing\cite{Wang2019,Burnett2019,Bylander2011}. We expect additional pure dephasing to manifest in a similar way qualitatively reducing gate fidelity further.

The left panel of figure \ref{fig:eVST1} shows the best fidelity that can be obtained for different values of the error and T\textsubscript{1}, while the right panel shows what is the value of the gate time which performs the best. The dark blue band at the bottom of the figure, is the gate duration of a non-robust pulse at 60 ns that is optimized numerically and reaches unit fidelity. The dashed red line represent the value of the control amplitude error at which robustness brings a benefit. It is clear that for a non-robust pulse the fastest gate which reaches unit fidelity is the most performant one, and will always performs best when there is no error, robust pulses do not compete in this regime. Moreover, the dashed line can be understood to come from comparing the Taylor expansion of the fidelity (see Appendix \ref{AP:error}) under both effect, $1-F = O(\exp(T/T_1)-1 = O(-T/T_1)$ and $1-F = O(\eta^2)$, and therefore the dashed line can be fitted to $\eta=k\sqrt{T/T_1}$ and match excellently.\\
The pulses which are robust for a control amplitude error are around twice as long as the non-robust pulses.
If the experimentally limited control amplitudes could be made greater then the duration of the pulses could also be shorter. This would change the figure \ref{fig:eVST1} by squeezing the plot horizontally to the left.\\ As the lifetime of qubits increases, the sensitivity to a control amplitude error increases as well.

The contour lines indicate what contribution governs the two mechanisms of fidelity loss. For example, when the contour lines are diagonal, the fidelity losses due to the error are of a similar order to the one by T\textsubscript{1}. It can be seen that for greater error values than the dashed lines the fidelities become insensitive to the error (the contour line is vertical). 
For the $T_1$ values of this system, the robustness is competitive from around $3\%$ which corresponds qualitatively to what is seen on the Figure \ref{fig:RB-timeStab}.

\subsection{Experimental verification}
\label{sec:experimental}
\begin{figure}[h!]
    \centering
    \includegraphics[width=0.6\linewidth]{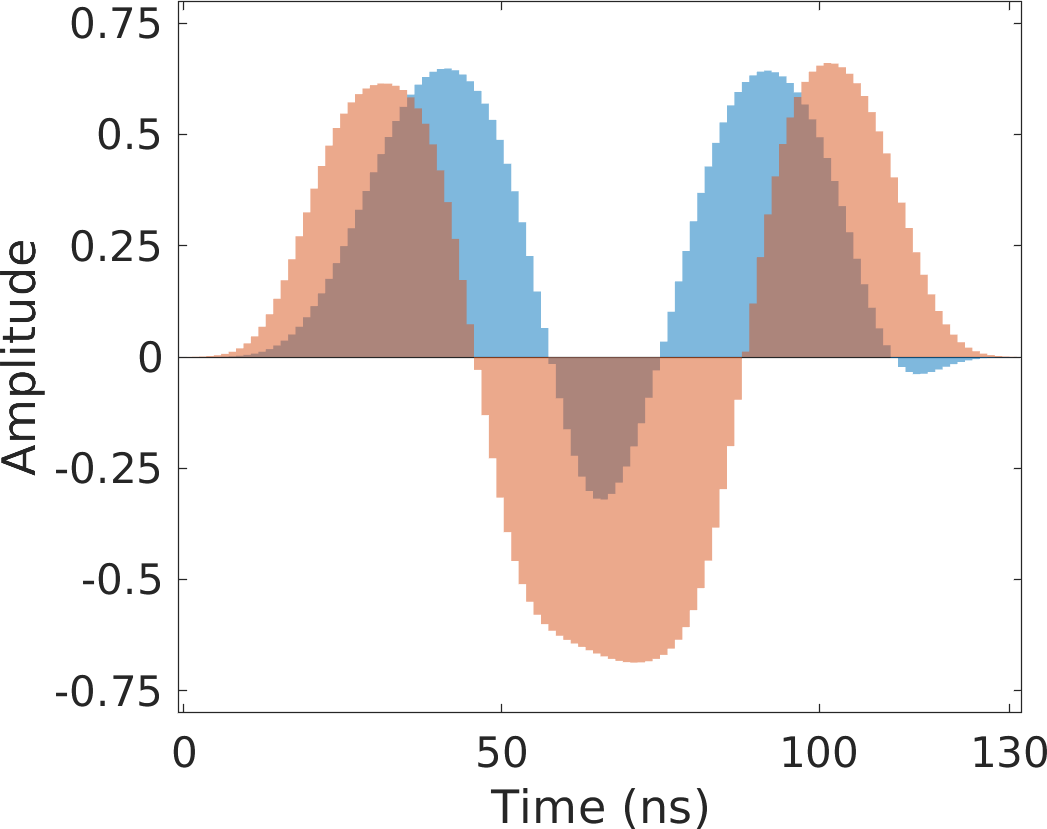}
    \caption{Robust control used for experimental verification. There are 25 control variables per channel, in blue is the $\eeps^x$ channel, and in orange is the  $\eeps^y$ channel.}
    \label{fig:control}
\end{figure}

%The controls are finally optimized in a polar coordinate system where the radius channel has fewer control variables than the angle channel. This method proves itself harder to find a high-fidelity solution in the robust case
%In order to maintain a high-fidelity solution, solutions found in the first part of this paper are re-used to generate a high-fidelity under constraints that are unfavorable to the optimization.
%\begin{equation}
   % min_{c_2} \sum(|CS(c_1(t))-CS(T(c_2))|)
%\end{equation}
%Firstly, a previously found pulse is used as a starting parameter under these new constraints.
A control pulse for a $X_{\pi/2}$ gate, with a duration of 130ns, see Fig.~\ref{fig:control}, was chosen for experimental verification. It was perturbed and then re-optimized three times to increase the infidelity to a value of 1e-5. 

The pulse's performance was measured using randomized benchmarking on the IBM Armonk qubit, where a random sequence of Clifford gates $g_1...g_L=I_2$ are concatenated on the qubit.
We generated six random sequences per Clifford string of length L with 1024 shots.
The generating set $G={ Z_{\pm\pi}, Z_{\pm\pi/2},X_{\pi/2}}$ can be used to express the Clifford group, on average one physical gate and 2.29 total gates, including virtual ones will be needed to generate an element of the Clifford group. This decomposition uses fewer gates than the one used for generating the entirety SU(2). The universal gate decomposition takes two physical gates and three virtual ones.  
All the Z-gates are done virtually by changing the phase of the subsequent control signals. These virtual gates are considered to have much higher fidelity than the physical ones. The error per gate was found to be ~3e-4 (EPG), showing that
the gates have good robustness against control error.
\begin{figure}[h!]
    \centering
    \includegraphics[width=\linewidth]{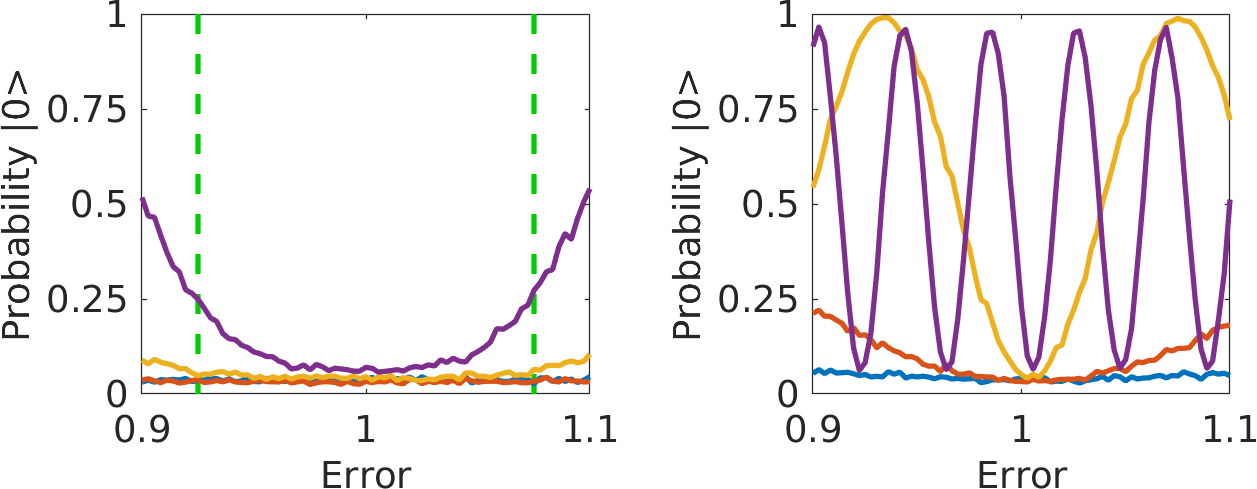}
    \caption{Experimental comparison between robustly optimized $X_{\pi/2}$ gate (right panel), and a half-DRAG pulse (left panel).
    The pulses are repeated 2, 6, 30, and 102 times indicated by the blue, orange, yellow, and purple lines. The dashed green line is the error range for which the robust pulse was optimized.}
    \label{fig:postCal}
    %SHOULD IT BE T_1/T_g
\end{figure}
We checked the robustness against an artificial perturbation.
The pulses are tuned experimentally by repeating the $X_{\pi/2}^N$ gate while varying the a) amplitude of the pulse b) the relative amplitude between the two control channels. The gates were found to have a phase error, this was measured by doing an amplified phase error (APE) sequence, $X_{\pi/2}(X_{\pi/2}X_{\pi/2}^\dagger)^NX_{\pi/2}$  \cite{Lucero2010,Kelly2010},  and corrected by adding a phase shift to the pulse.
The robustness of the pulses can be seen in the range of figure \ref{fig:postCal}.
%\ref{fig:RB-decay}.
\subsection{Tuning}
In order to implement a robust solution in hardware, the Hamiltonian \ref{eq:hami} is defined by only a few parameters, which might not correspond exactly to the signal the qubit will receive.
These imperfections can be calibrated by fine-tuning the amplitudes of the control channel: especially by amplifying amplitude errors and by amplifying the phase error using an APE scheme. A phase error occurs on the gate, $X'_{\pi/2}=Z(\epsilon)X_{\pi/2}Z(\epsilon)$, which is tackled using a phase virtual-Z gate in between controls (see appendix \ref{sec:fineTune}). 
Furthermore, different errors might interfere such that the value measured for one might be different from its actual value, for example, a phase error will change the value of the amplitude of the pulse during the tuning. 
 
The gate was verified against an artificial error $c'(t)=d\cdot c(t)$ to simulate an error on the amplitude of the control like the one optimized for(see Fig.~\ref{fig:RB-timeStab}). The factor $d$ was in the range of $d\in[0.9,1.1]$. The error was found to be no worse than $3e-4$.
The fidelity is also relatively stable in a three day period, also maintaining robustness even at the end of the experiments.

\section{Conclusions}
Robust optimal control can be used to generate implementable pulses that have high fidelity in experiments. Although these solutions are significantly longer than non-robust solutions, by a factor of around 2, they can be competitive if the control amplitude is large enough. As the lifetime of qubits increases, the sensitivity to control amplitude errors also increases.\\

Furthermore, these solutions need to be tuned in hardware to extract the fidelity predicted by the theoretical results. Since the coherent error on these pulses is smaller than the measurement error, error amplifying schemes become a useful and necessary tool to tune the pulses for achieving high fidelity. This tuning was done by varying only a few parameters, such as the relative control amplitude and an additional virtual phase gate. The performance of the pulses was verified using randomized benchmarking while introducing an artificial control amplitude error, and was found to achieve both robustness and high fidelity.\\

The control landscape for these robust solutions changes with the error. At low control amplitude error, the controls become trapped in sub-optimal maxima, and their error profiles match those of non-robust solutions. Large control amplitude errors significantly change the landscape, making robust solutions easier to find. Between these two zones, there is a transition between non-robust and robust solutions.\\

Robust optimization is prone to getting trapped in sub-optimal solutions, but counter-intuitively, increasing the error may facilitate finding a better solution. Solutions can also be trapped when the gate time is close to the minimum time required for robust solutions to be found. A small perturbation in the control can nudge it out of this region, allowing a higher fidelity solution to be found.

\section{Grant and funding}
This work is supported by the UK Hub in Quantum Computing and Simulation, part of the UK National Quantum Technologies Programme with funding from UKRI EPSRC grant EP/T001062/1, and the EPSRC strategic equipment grant no. EP/L02263X/1. All the codes and simulation data of this paper are available without restrictions.

\clearpage
\appendix

\section{Fine tuning of a few parameters of the control}
\label{sec:fineTune}
\begin{figure}[h!]
    \centering
    \includegraphics[width=0.6\linewidth]{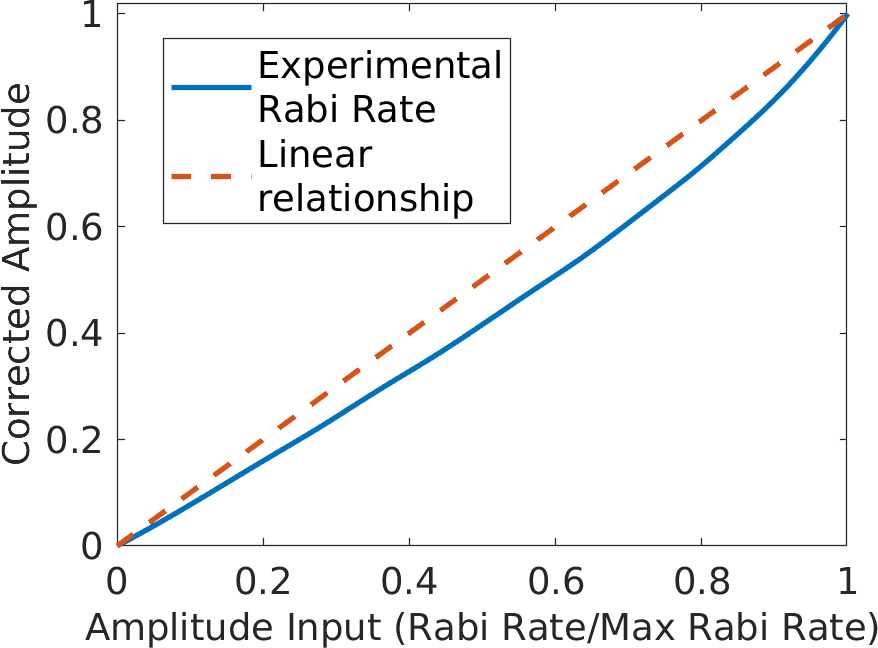}
    \caption{Inverse transfer function: mapping the desired control input to the output amplitude. The dashed line represents a linear mapping from the maximum obtainable Rabi-rate.}
    \label{fig:amplitude_Transfer}
    %SHOULD IT BE T_1/T_g
\end{figure}

The Rabi rate is measured at different input amplitudes using flattop Gaussian functions with fixed rise and fall time using different duration gates. The rise and fall time lead to an added constant phase on the measured Rabi oscillations. The relationship between the amplitude-input and Rabi rate is not linear and differs significantly when the relative amplitude is greater than 0.4 (see Fig.~\ref{fig:amplitude_Transfer}), the transfer function is made continuous using cubic splines.

The control amplitude can be corrected post-optimization as there is always a mapping:
\begin{equation}
    c(t)=T_a\bigg[\sqrt{\eeps_x(t)^2+\eeps_y(t)^2}\bigg]\frac{(\eeps_x(t)+i\eeps_y(t))}{\sqrt{\eeps_x(t)^2+\eeps_y(t)^2}},
\end{equation}
where $T_a$ is the inverse amplitude transfer function.

\begin{figure}[h!]
    \includegraphics[width=0.48\linewidth]{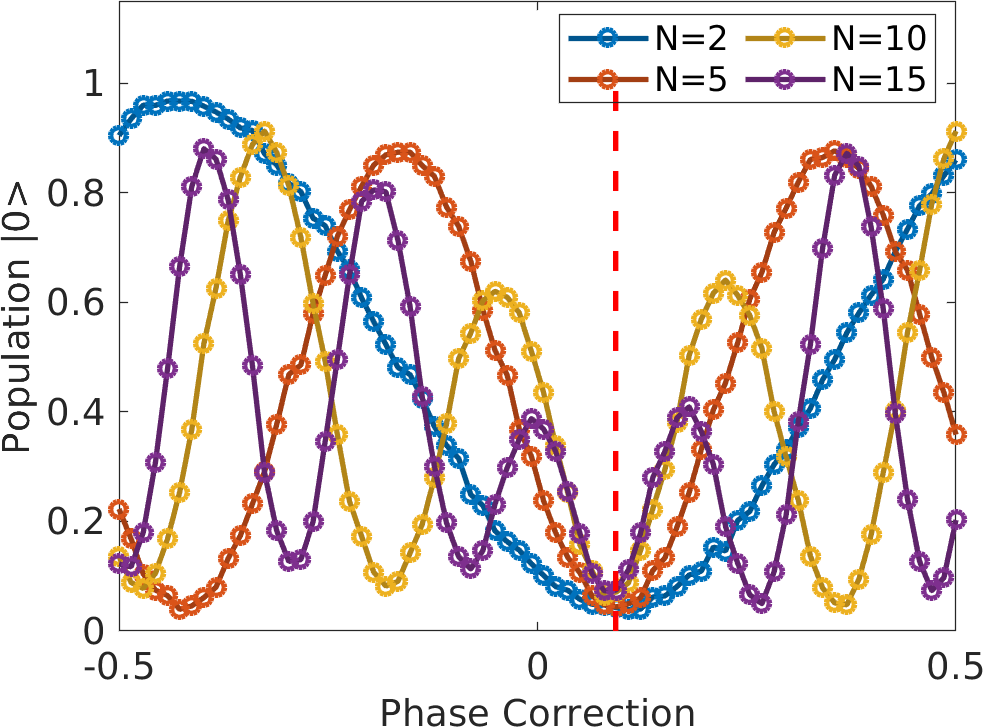}
    \includegraphics[width=0.435\linewidth]{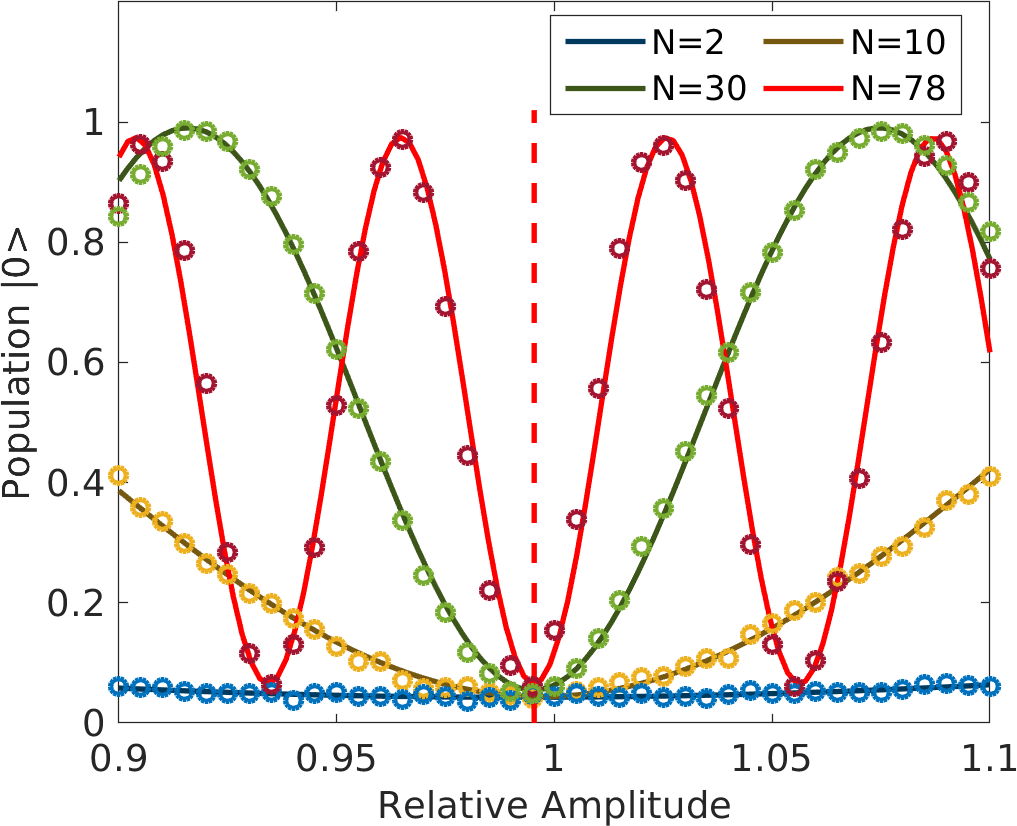}
    \caption{Calibration of a $X_\frac{\pi}{2}$ gate. Left panel: Calibrating the phase error using an APE sequence, $X_{\frac{\pi}{2}}\left(X_{\frac{\pi}{2}}X_{-\frac{\pi}{2}}\right)^NX_{\frac{\pi}{2}}$, and the dashed red line is the phase correction which minimizes the sum of the curves.
    Right panel, calibration of the relative amplitude by repeating an $X_\frac{\pi}{2}$ and the dashed red line is the amplitude correction which minimizes the sum of the curves}
    %\label{fig:amplitude_corrections}
    %\label{fig:phase_corrections}
    \label{fig:tunings}
\end{figure}

When calibrating the pulses, a significant phase error (see the right panel of Fig.~\ref{fig:tunings}) was observed. Therefore, to calibrate it out, we swept a phase correction while carrying out APE sequences, which amplify the phase error (at small phase error) linearly and found the value which minimizes the population in the ground state.
These sequences are repeated at different values such that the correction value corresponds to the minimum on the totality of the curves.

Another pulse deformation is the channel-dependent error, $a$:
\begin{equation}
    c(t)=\eeps_x(t)+ia\eeps_y(t)
\end{equation}
The parameter $a$ is swept in an amplitude error amplification of $X_\pi/2$ experiments to infer its correct value. The value found is $a=0.995$ which is very close to having no such deformation (see Fig.~\ref{fig:tunings}).

\section{Difficulty of finding solutions}
\label{sec:Hrdness}
The probability of finding a high-fidelity solution depending on the value of $\eta$ is, however, not simple:  we consider the probability of the solutions to have an infidelity within a certain 'distance' to the best fidelity obtained for a specific error range, according to this formula: $Q(F_\eta,d)=Pr(\log_{10}(1-F_{\eta,max})-\log_{10}(1-F_\eta)<d)$. 

\begin{figure}[h!]
    \includegraphics[width=0.48\linewidth]{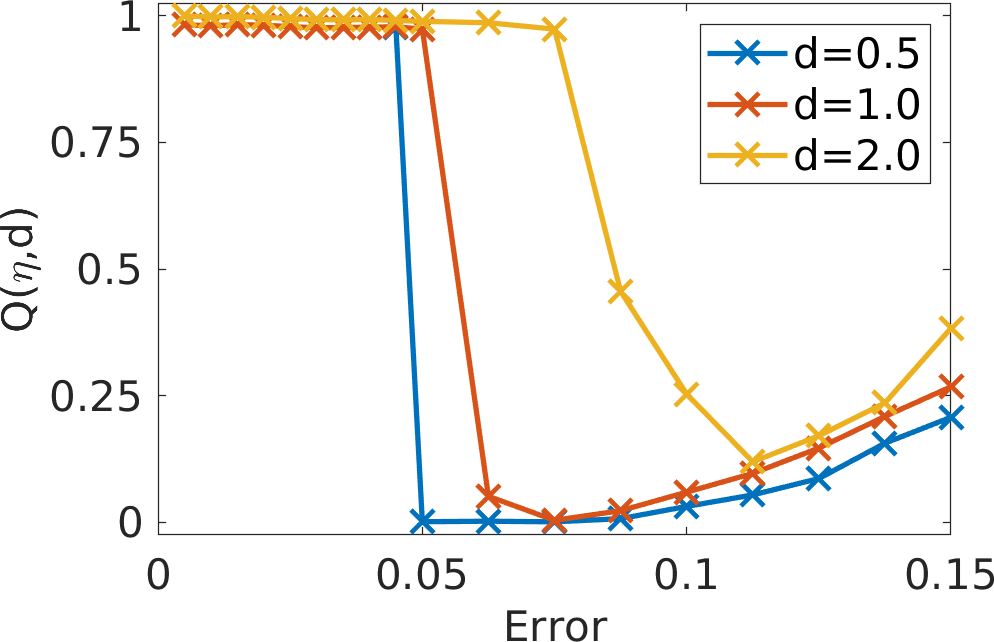}
    %\label{fig:distanceFR}
    \quad
    \centering
    \includegraphics[width=0.4\linewidth]{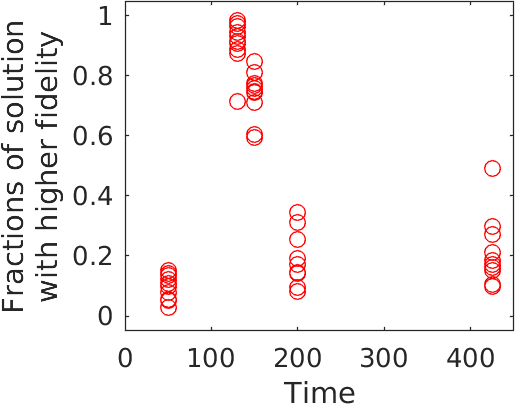}
    \caption{The difficulty of finding the best solution for different errors and gate times. Left panel: Fraction of controls within a log infidelity difference of -0.5(Blue), -1 (Red), and -2 (yellow) to the maximum fidelity found. The Hamiltonian has 25 control variables per channel, and the gate time is T=130ns. Right panel: Fractions of controls that have a higher fidelity after being perturbed and then re-optimized once.}
    %\label{fig:fractionsOpt}
    \label{fig:fractions}
\end{figure}

When $\eta<0.05$, which is the region where the optimizer does not find the control with higher fidelity and not a robust solution (see Fig.~\ref{fig:fractions}), the fidelity across the error range has a similar shape to the DRAG fidelity curve, but it has a slightly higher fidelity at the extrema and is therefore insignificantly more robust than DRAG (see Fig.~\ref{fig:robustBenefit}). In this first range, almost all controls have an infidelity within 0.5 from the maximum infidelity, however, after $0.05<\eta<0.1$, there is a sharp change in the probability of finding a solution of the same order as the maximum one, the probability swings from exceeding 95\% to below 1\%. Finding controls with similarly high fidelity as the maximum fidelity found is rare. As a consequence of this, a large number of starting points are needed in order to find this optimal solution.
When the gate time $T$ is close to a value at which a robust solution can be found (the minimal robust gate time), not only is it hard to find a high fidelity solution (see the left panel of Fig.~\ref{fig:fractions} ,\ref{fig:robustBenefit} and appendix \ref{AP:avvswc}) the optimization is also considerably more likely to get 'trapped' on a sub-optimal solution (similar statements apply for non-robust optimization \cite{Larocca2018}). The difficulty in finding fast gates is highly relevant when the evolution is non-unitary, as the gate time will limit the maximum fidelity.

\section{Perturbation of the control}
\label{sec:pert}

The optimization may be trapped on sub-optimal maxima but may be in the vicinity of a better solution. Therefore, perturbations may move the control far enough from the trap and enable finding a better solution.
In order for the optimizer to accept a perturbed starting points, they must satisfy these constraints:
\begin{align}
    c_i+p_i< c_{max}, &&  c_i+p_i>c_{min}\\
    |c_i+p_i-c_{i+1}|<slew, && |c_i+p_i-c_{i-1}|<slew\\
    p_i<p_{max} && p_i>p_{min}
\end{align}
These constraints are convex, and therefore the feasible values for $p_i$ will be in the range $p_{i,min}\leq p_i\leq p_{i,max}$.
The space that fulfills the conditions is called the feasibility area. The control starts in the feasible area and must remain there after the perturbation.

The perturbation is taken uniformly in the range of feasible values for the control variable perturbation.    
The perturbation's upper and lower bounds of a control can be nearly equal. Therefore the perturbation benefits from repeated cycles of perturbations (before re-optimization) in order to increase the volume of the perturbation space.
\begin{figure}[h]
    \centering
    \includegraphics[width=0.65\linewidth]{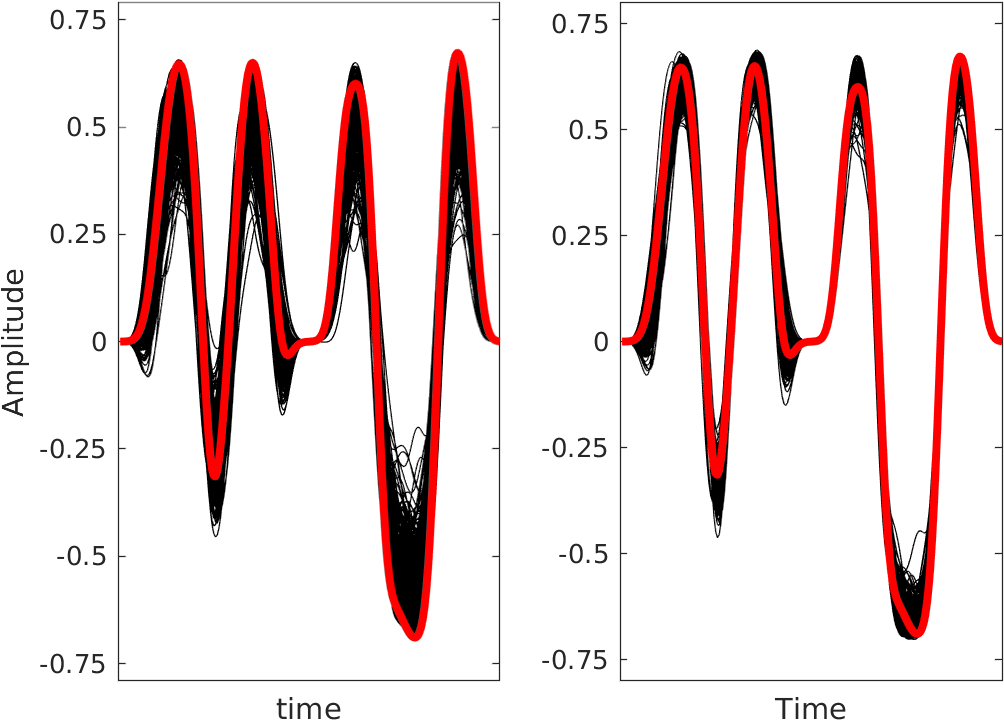}
    \caption{Perturbation of the control in order to reach a higher fidelity. In the left panel in red, the original control and in black, the perturbed controls.In the left panel, controls after having been optimized using the controls in the right panel as starting points. In both subplots the control channels are concatenated with $\eeps_x$ then $\eeps_y$.}
    \label{fig:perturbationControl}
    %SHOULD IT BE T_1/T_g
\end{figure}

The controls for a high-fidelity solution could be perturbed by $~1\%$ per time bin until the average infidelity: $1-F_{u}$ decreased by a factor of 10,  where $F_u$ is the fidelity of the unperturbed control. In some cases, controls with slightly higher fidelity were found, but the computational cost was prohibitive to use as a general optimization strategy.\\
\begin{figure}[h!]
    \centering
    \includegraphics[width=0.7\linewidth]{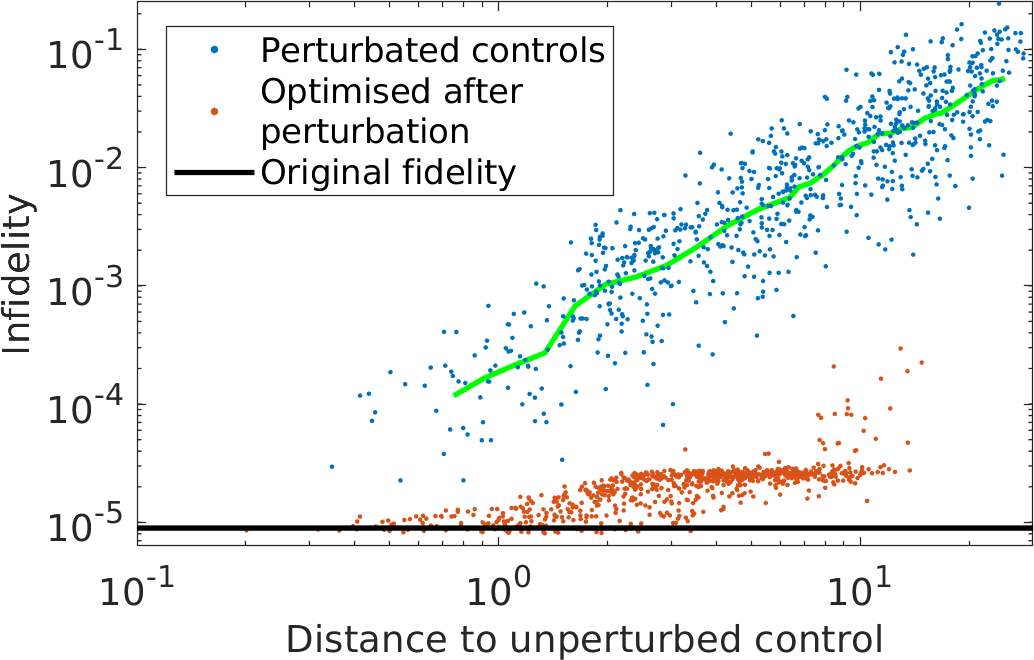}
    \caption{Infidelity in the vicinity of the optimal control and convergence after two cycles of perturbation and re-optimization. The control vector was perturbed 800 times, and the Manhattan norm between the bins of the resulting time signal was calculated. The green line is the rolling average of the infidelity of the perturbed control. }
    \label{fig:pert}
\end{figure}

In landscapes with a higher error, $\eta>0.1$, perturbing the controls did not result in a lower or higher fidelity despite the controls being different, indicating the abundance of solutions with similarly high fidelities. The perturbed controls after the optimization tend to return closer to the original control, but their fidelities do not return to the original maxima and get trapped on a lower fidelity(see Fig.~\ref{fig:pert}). In fact, in that figure it can also be seen that a lot of solutions get trapped at on lower fidelities around $log_{10}(1-F)=-4.5$ .
%The control landscape for robust optimization has high dimensionality making it difficult to visualize, in order to quantify the differences between controls and to extract the landscape's relevant features, we will perturb the controls and compute the distance to the original control in order to analyze the solutions obtained.
%The controls obtained after the optimization were analyzed using Matlab's DBSCAN with euclidean distance to cluster the controls, and at different values of $\eta$, the behavior varies slightly: very few clusters are found. When clusters were found, they had only a few members, and the overwhelming majority of controls did not form clusters, indicating that the landscape is rich and there are an abundance of solution with similar fidelity.

\iffalse
\subsection{Randomized Benchmarking}
\begin{figure}[h]
    \centering
    %\includegraphics[width=0.75\linewidth]{RB-sample.png}
    \caption{\note{Check}Example of randomized benchmarking, with 20 experiments with two repetitions per Clifford string length.}
    \label{fig:RB-decay}
    %SHOULD IT BE T_1/T_g
\end{figure}
\fi

\section{Comparison with composite pulse sequences}
\label{AP:bb1}
\begin{figure}[h]
    \centering
    \includegraphics[width=0.75\linewidth]{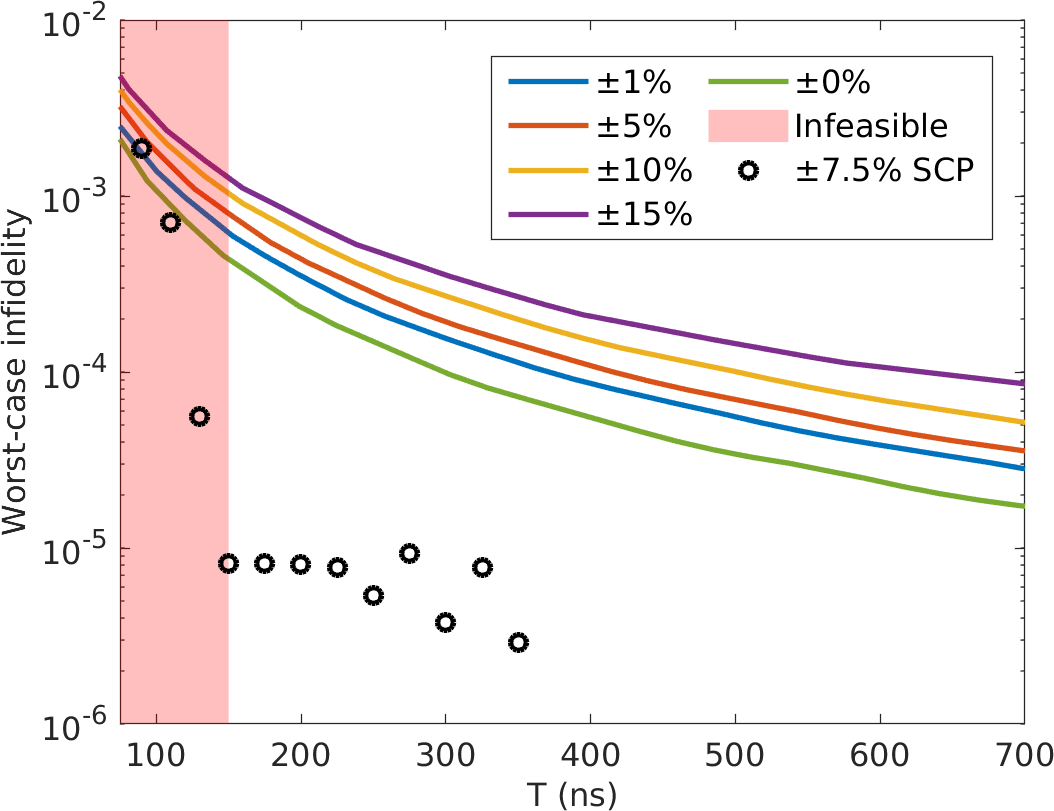}
    \caption{The black circles is the infidelity found using SCP.
    The infidelity of the BB1 for different curves are noisy, so the lower part of the bounding curves (solid lines) at different times are plotted.
    The red region is the region where the composite pulse sequence would have to use a control amplitude higher than the maximum one available. }
    \label{fig:BB1-Fid}
    %SHOULD IT BE T_1/T_g
\end{figure}

Composite pulse sequences, use a set of rotations for different polar and azimuthal angles of rotation. For instance, a composite pulse sequence for BB1 \cite{Wimperis1994} would be:
\begin{equation}
    BB1(\theta)=180_{\phi_1}360{\phi_2}180_{\phi_1}\theta_0
\end{equation}
Where $\theta=90$ for a $\pi/4$ rotation and the $\phi_i$ are the phases in the signal, corresponding effectively to the driving between $\sigma_x$ and $\sigma_y$. Immediately, it is clear that this composite sequence needs to last at least as long as if it was driven without rise and fall time, at the maximal amplitude. Therefore, the minimal time a composite pulse sequence can have is $T_{BB1}=2\pi\lambda_1^{-1}(180+360+180+\theta)/360$, and for a $X_\frac{\pi}{2}$ gate the minimal time for a BB1 sequence is 150 ns (see figure \ref{fig:BB1-Fid}), which is longer than the pulses used in the text. \\
Unfortunately, when these pulses are modulated on a square envelope they generate spectral components at the value of the anharmonicity, this excites higher order transition and therefore lowers the fidelity in the case of a transmon.
While this problem can be remedied by using pulse shaping techniques, the gate time is increased even more, making the BB1 pulse less competitive when it comes to incoherent effects.

The main issue is that the BB1 pulse sequence is not designed to handle multi-level systems, and in a naive implementation of BB1, one that uses square pulses, the worst-case fidelity is significantly worse and the gate time is also longer. Given the minimum time required to do a BB1 pulse is around 150ns and has a worst-case fidelity around 99.9\%, given a 5\% error. At similar gate times, BB1 perform two orders of magnitude worse than the robust pulses devised in previously.
The non-robust performance is also worse than a simpler shaped pulse. BB1 pulses need significantly longer gate time to reach high fidelity (see figure \ref{fig:BB1-Fid}).

\section{Optimizing for the worst case fidelity and the average case fidelity}
\label{AP:avvswc}
\begin{figure*}
    \centering
    \includegraphics[width=0.9\linewidth]{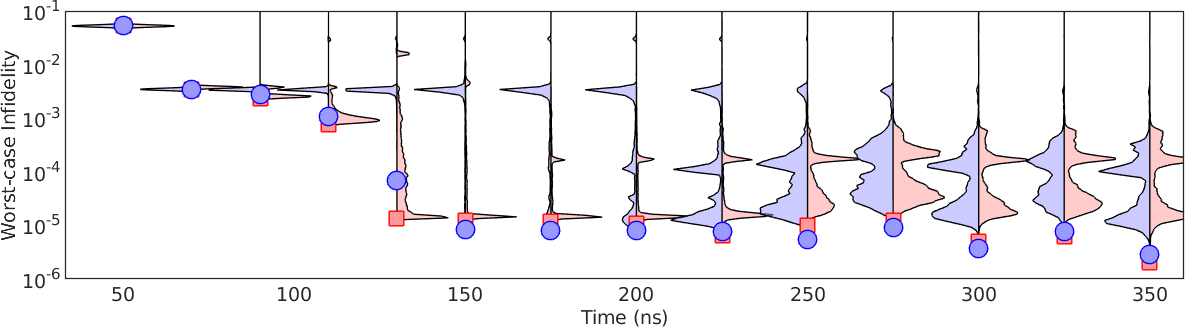}
    \includegraphics[width=0.9\linewidth]{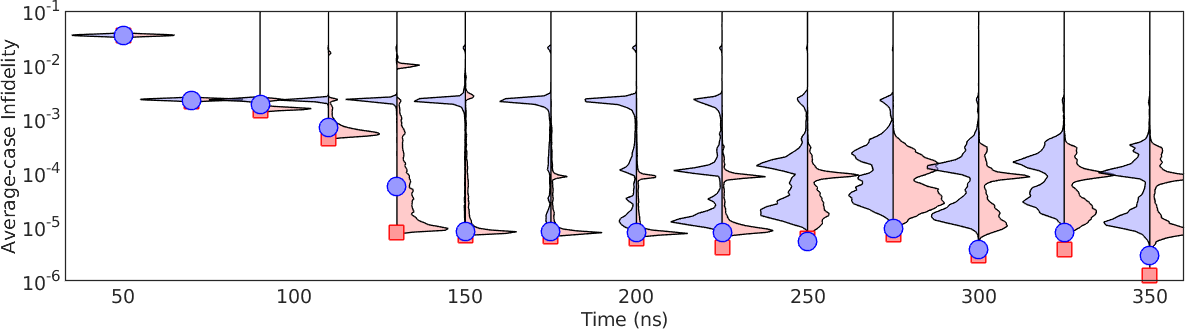}
    \label{fig:avVSwc}
    \caption{Comparing optimization strategies. In blue, infidelity distribution optimizing the worst-case, and in red, infidelity distribution optimizing the average-case. The amplitude of each individual curve is a probability density over the results of the optimization, the markers in each of the plots respective colors corresponds to the best result. The top panel shows the worst-case fidelity, of each optimization strategy. The bottom panel shows the average case.
    The optimization is carried out for 5000 starting points, 20 control parameters on per channel and for one cycle of optimization.}
\end{figure*}

The optimization convergence and trapping is different when optimizing for the average case fidelity.
It can be seen from the top panel of figure \ref{fig:avVSwc} that optimizing for the worst-case fidelity gives in general  slightly better worst-case fidelity but the performance is comparable to optimizing the average case fidelity, except at $T=130$. In those cases, it can be seen that the optimization gets trapped on lower fidelity solution, namely at around $F=99.5\%$, which correspond to the fidelity profile of a non-robust pulse.\\

\begin{figure}
    \centering
    \includegraphics[width=0.75\linewidth]{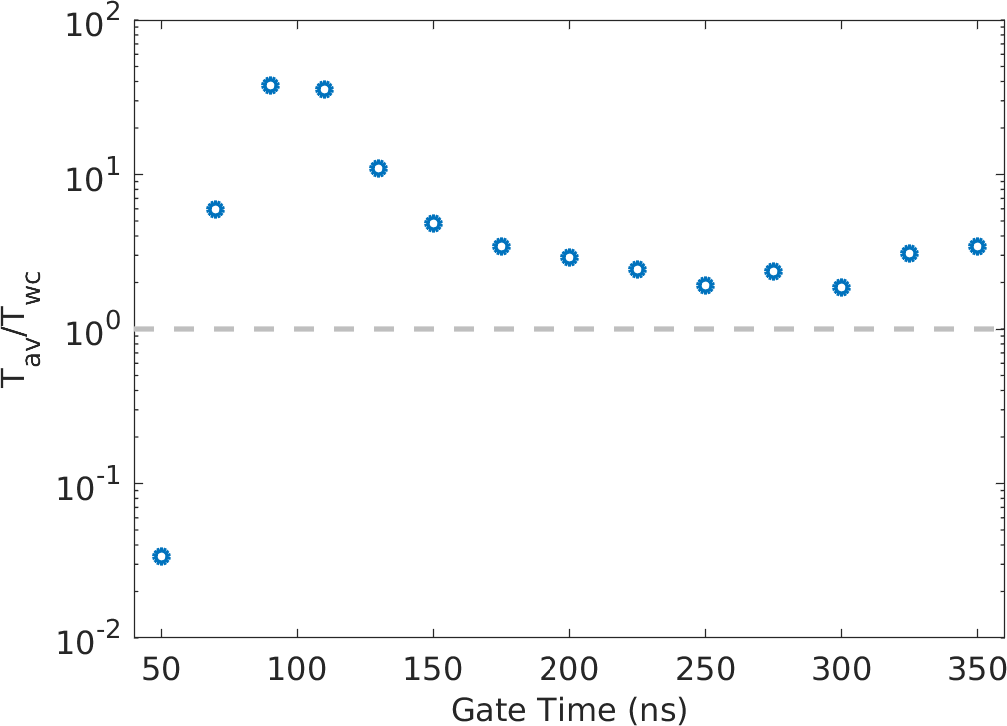}
    \caption{Ratio of the optimization time for the average case $T_{av}$ and the optimization time for the worst case optimization, $T_{wc}$ for different gate times.}
    \label{fig:timeopt}
\end{figure}

When using the same optimization parameters, the duration of the optimization of the average-case is much larger than the worst-case (see figure \ref{fig:timeopt}).
In fact the median optimization time is more than 3 times greater, and in the worst cases the time may be $\sim$30 times greater.

\section{Taylor expansion of the fidelity with a detuning and amplitude error for a qubit}
\label{AP:error}
The fidelity of a non-robust operation will degrade in the presence of an amplitude ($\epsilon$) or detuning error ($\delta$).
Let us consider a two-level qubit Hamiltonian:
\begin{equation}
    H=\delta \sigma_z + (1+\epsilon)\frac{c(t)}{2}\sigma_x,
\end{equation}
If we are to implement a $X_\frac{\pi}{2}$ on resonance, the control in an error-free ($\delta=\epsilon=0$) simple trajectory must satisfy
\begin{equation}
    \int_0^{T_g}c(t)dt=\frac{\pi}{2}.
\end{equation}
For the sake simplicity, let's consider that the control is constant, $c(t)=\frac{\pi}{2T_g}=\alpha$.
We can calculate the error contributions by calculating the time evolution operator which in this case takes the form:
\begin{gather}
    U(t)=\cos(\theta T_g)I - i\sin(\theta T_g)\left(\frac{\delta}{\theta}\sigma_z + \frac{\alpha}{\theta}\sigma_x\right)\\
    \text{where: }\theta=\sqrt{\delta^2+\alpha^2}
\end{gather}
In this case the fidelity, can be calculated analytically as:
\begin{equation}
	F=\frac{1}{4}\left|\mathrm{Tr}(U(T_g)X^\dagger_\frac{\pi}{2})\right |^2,
\end{equation}
and if this formula is expanded by considering a perturbation in detuning and the error at $\epsilon=0$, than to second order:
\begin{equation}
    F=1-\frac{8\,T^2\,\delta^2}{\pi ^2}+O(\delta^4).
\end{equation}
While if considering a perturbation on the control amplitude:
\begin{equation}
    F=1-\frac{\pi ^2\,\epsilon^2}{16}+O(\epsilon^4).
\end{equation}
These two errors affect the fidelity differently, in the case of an error on the detuning, the fidelity loss during the gate is proportional to the square of the product of the gate time and detuning, while for the amplitude error there is no gate time dependence.\\
This provides an immediate strategy to reduce the error due to a detuning error namely reducing the gate time.

Furthermore, when the drive on a multi-level has little spectral weight on the $\omega_{12}$ transition, the infidelity contribution will correspond to that of a qubit.

\section{Frequency robust and doubly robust pulses}
\label{AP:freqRob}
\begin{figure}
    \centering
    \includegraphics[width=0.75\linewidth]{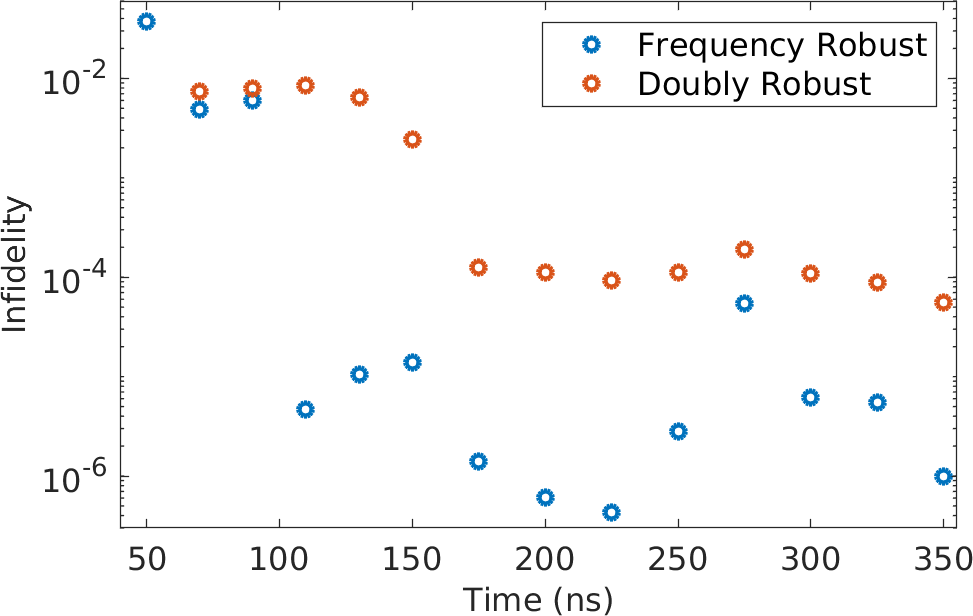}
    \caption{Infidelity of frequency robust and doubly robust. The error optimized for is 500 KHz for the frequency error and $7.5\%$ for the control amplitude error.}
    \label{fig:freqInfid}
\end{figure}

The optimizations techniques can also be used to make pulses robust to both a frequency error. As well as robust to both control amplitude error and frequency error.\\
It can be seen in figure \ref{fig:freqInfid} that robust pulses to frequency error can can be found similarly fast as amplitude robust pulses, and the infidelity obtained is lower than 1e-5. The behavior changes dramatically when the gate time is long enough to allow for a robust pulse.
The time required to obtain pulses which are doubly robust with high is significantly higher, around 175ns.
\clearpage
\bibliographystyle{ieeetr}
%\bibliography{source}
%%% biblo

\providecommand{\noopsort}[1]{}\providecommand{\singleletter}[1]{#1}%

\end{document}